\documentstyle[aps,epsfig,eqsecnum]{revtex}

\newcommand{\bc}{\begin{center}}
\newcommand{\ec}{\end{center}}
\newcommand{\bg}{\begin{equation}}
\newcommand{\ed}{\end{equation}}

\newcommand{\kt}{k_{\rm B}T}
\newcommand{\lb}{l_{\rm B}}
\newcommand{\di}{\frac{\tilde N - \tilde M}p}
\newcommand{\rr}{{2\tilde m\over \tilde N}}

\newcommand{\qe}{Q_{\rm e}}

\begin{document}
\draft
\title{Stretching necklaces}
\author{T.A. Vilgis, A. Johner, J. F. Joanny} 
\address{Laboratoire
Europ\'een Associ\'e, Institut
Charles Sadron, 6, rue  Boussingault,\\ 
F- 67083 Strasbourg, Cedex,\\
France} 
\date{\today}
\maketitle

\begin{abstract} 
Polyelectrolytes in poor solvents show a necklace
structure where collapsed polymer
pearls are linked to stretched strings. In the present paper the 
elasticity of such chains is studied in detail. Different 
deformation regimes 
are addressed. The first is the continuous regime, where many
pearls are present. A continuous force extension relation ship is 
calculated. The main contribution comes from the tension balance
and the electrostatic repulsion of consecutive pearls.
The main correction term stems from the finite size of the 
pearls, which monitors their surface energy. For a finite amount
of pearls discontinuous stretching is predicted. Finally
counterion effects are discussed qualitatively.

\end{abstract}


\section{Introduction}
Recently it has become possible to manipulate single polymer chains and to
experimentally study their elastic response \cite{chatellier:98,wilder:98}.
Our aim here is to describe theoretically the single chain elasticity in the
specific case of polyelectrolytes in poor solvents, typically in water, a poor
solvent for most organic polyelectrolyte backbones. Beside the direct
experimental measurements of single chain elasticity, there are at least two
other motivations for this study: the possible relation with drag reduction
\cite{degennes:86,mumick:94}, and the relation to the elastic behavior of more
complex structures such as polyelectrolyte gels.

Charged polymers in solution, polyampholytes and polyelectrolytes, are known
to reduce drag very efficiently in turbulent flows.  This property is shared,
to some extent, by neutral polymers.  For neutral polymers, drag reduction has
sometimes been explained by the dissipation associated to the stretch / coil
transition where the polymer chains undergo a discontinuous elongation. One
would also anticipate a large dissipation for a collapsed polymer globule in a
poor solvent: even for fairly small variations of the stress there is a
significant hysteresis loop in the globule/stretched chain transition.  Of
course, the concentration of polymer globules in a poor solvent must be
extremely low in order to avoid phase separation. It has been suggested
recently that polyelectrolytes in poor solvents may exhibit a pearl-necklace
structure where collapsed globules, the pearls, are connected by stretched
polymer strings. One may look at this system as a way of solubilizing
(connected) dense polymer globules at finite concentrations, these structures
could then be good candidates to enhance drag reduction.

If the necklaces comprise many pearls (which, we think, might be difficult to
achieve for single chains in the accessible experimental range of parameters),
thermal fluctuations smear out the single pearl features and the
force/elongation curve is continuous. The continuous elasticity of these
necklaces may be used as a starting point to model the elasticity of more
complex, systems : polyelectrolyte gels will be described in a separate paper.

The behavior of polyelectrolyte chains in a Theta solvent has been discussed
thoroughly theoretically and it is fairly well understood (we will not need in
this paper to consider the bond angle correlations and the persistence length
that remain controversial issues). We consider a chain of $N$ monomers of size
$b$, with a fraction $f$ of charged monomers. The repulsive Coulombic
interaction between charged monomers is written as $V({\bf r}) = \kt \lb /r$,
$\lb = e^2/(4\pi \epsilon k_{\rm B}T)$ being the Bjerrum length where
$\epsilon$ is the dielectric constant of the solvent and $e$ the elementary
charge.  The chain radius can be estimated by a balance between entropy and
electrostatic energy that is equivalent to the electrostatic blob model
\cite{degennes:76}: the chain may be viewed as a stretched string of Gaussian
isotropic blobs each containing $(l_B f^2/b)^{-2/3}$ monomers and of radius
$(l_B f^2/b)^{-1/3}b$; the overall chain length is then $R = b
N\left(\frac{l_{\rm B}f^2}{b}\right)^{1/3}$.  The electrostatic interaction is
relevant for high enough charge fraction i.e., $f>N^{-3/4}$, otherwise the
chain remains essentially Gaussian isotropic (comprising less than one blob).

If the chain backbone is in a poor solvent (which is often the case in
experiments), there is a competition between surface tension effects tending
to minimize the polymer/water(solvent) contact area and electrostatics tending
to maximize the overall chain size. In the following, the poor solvent
conditions are characterized by a negative excluded volume $v = - \tau b^3$,
where $\tau = (\theta -T)/\theta$ is the relative distance to the compensation
temperature $\theta$. In the absence of electrostatic interactions, the chain
collapses into a dense globule that may be viewed as a small region of dense
polymer phase at co-existence with free solvent; the finite size of the
globule is associated with an extra energy penalty due to the polymer-water
surface tension. The balance of the osmotic pressure between the dense and
dilute phases (of almost vanishing concentration) yields the concentration
inside the globule $c = \tau / b^3$. Alternatively, the dense phase may be
described by a close packing of thermal blobs of size $\xi_t = b/\tau$,
containing $g=1/\tau^2$ monomers. A globule of $N$ monomers has then a radius
$R = b \left( N/\tau \right)^{1/3}$. The corresponding surface tension is
$\gamma = k_{\rm B}T/\xi_t^2 \propto \tau^2/b^2$. For a charged chain, the
balance between surface tension and electrostatics was first considered by
Khokhlov assuming a cylindrical globular shape\cite{khokhlov:80}. This yields
a highly elongated globule of length $L = N b (\lb f^2/b)^{2/3}/\tau$ and
radius $R_c = b (\lb f^2/b)^{-1/3}$ provided that the radius of the chain in
the absence of electrostatic interactions, in a spherical globular
conformation, is larger than $R_c$; for smaller chains, the electrostatic
interaction is only a perturbation. The elongated cylindrical globule can be
looked at as a stretched string of spherical globules of radius $R_c$.

Recently, it has been proposed by Kantor and Kardar \cite{kantor:94,kantor:95}
that the large spherical globule instability is rather similar to the Rayleigh
instability of charged liquid droplets. A large liquid droplet carrying an
electrostatic charge, breaks up into marginally stable smaller droplets for
which surface tension and electrostatic self-energy are of the same order of
magnitude. For a "liquid of thermal blobs" with individual size $\xi_t$, the
size of the marginal stable droplet (the pearl) is again $R_c$.  However, in
order to minimize the total free energy, the droplets must be infinitely far
apart. Due to the chain connectivity, the best separation that can be achieved
in the polyelectrolyte problem is to connect the pearls with stretched
polyelectrolyte strings. This leads to the pearl-necklace model introduced by
Dobrynin, Rubinstein and Obukhov \cite{dobrynin:96}. The string monomers are
then in equilibrium with the pearl monomers. An equilibrium between a
collapsed chain and a globule is also found when an external force is applied
on a globule. As shown by Halperin and Zhulina \cite{halperin:91.1}, in a
first approximation, (discarding finite size corrections) the strings must
have a radius equal to $\xi_t$ (the thermal correlation length in the pearls),
and a tension $\sigma=\frac{\kt}{\xi_t}$. The string tension is qualitatively
due to the repulsion between adjacent pearls $\kt \lb Q^2/l^2$ (a logarithmic
factor corresponding to pearls farther apart is omitted), where $Q$ is the
pearl charge and $l$ the length of a string. The string length is determined
by the force balance $\sigma =\kt \tau/b = \kt \lb Q^2/l^2$. It is easily
checked that, if the solvent is poor enough, most monomers belong to pearls
and that the stretched strings dominate the total length $L = (f^2\lb
/b)^{1/2} N b \tau^{-1/2}$.

\begin{figure}
\begin{center}
\begin{minipage}{12cm}
\centerline{{\epsfig{file=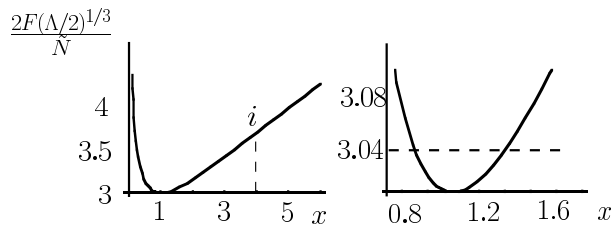,width=8cm}}} 
\vspace*{10pt}
\caption{\label{fig:necf1} 
{\footnotesize
Free energy per thermal blob of an assembly of  isolated droplets 
  comprising $x \Lambda/2$ thermal blobs each. The monodisperse assembly is
  locally stable for droplet sizes up to four times the preferred size
  $\Lambda/2$. For droplets larger than $1.4 x$ it is energetically favorable
  to split a droplet into two identical smaller droplets (see enhanced
  figure). Fluctuating between $x=0.7$ and $x=1.4$ only involves a fraction of
  the thermal energy per droplet for $\Lambda = 100$.  
}
}
\end{minipage}
\end{center}
\end{figure}

It is worthwhile to return to the case of an ordinary simple liquid, the
Rayleigh problem, where a large drop splits into non-interacting smaller drops
far apart. In order to remain close to the polyelectrolyte problem, we
consider a "formal" liquid of disconnected thermal blobs. The free energy of
an assembly of $\tilde N = N\tau^2$ blobs grouped into drops of $\tilde
m=m\tau^2$ blobs each is the sum of the electrostatic and interfacial energies
of the individual drops. It reads $F/\kt = 0.5\tilde N (\Lambda/2)^{-1/3}
\mbox{f}(x)$ where $ x = 2\tilde m/\Lambda$ is the number of blobs in the
marginal droplet. Note that $\Lambda =\tau^3/(\lb f^2/b)$ (hence $\Lambda>1$
is required). The function $\mbox{f}(x) = 2 x^{-1/3} + x^{2/3}$ is plotted in
Fig.\ref{fig:necf1}:
\begin{itemize}
\item It has a minimum at $x=1$, the preferred droplet size contains $\tilde m
  =\Lambda/2$ blobs.
\item It
changes curvature at $x=4$: an assembly of identical droplets smaller than
four times the
preferred size is locally stable against polydispersity.
\item For a size larger than about 1.4 times the preferred size,  it is
favorable to
split a droplet into two identical droplets.
\item The overall free energy $\mbox{f}(x)$ is rather flat and thermal
fluctuations are
anticipated  to become
important for mesoscopic systems. For an assembly of 1000 blobs (a
rather large number) and a preferred
droplet size of 100 blobs, the barrier against splitting all 10 droplets is
roughly $4\kt$ only.
The fluctuations become less important with 
decreasing solvent quality.
\end{itemize}

Though we are discussing here a liquid of disconnected blobs only, we may
anticipate that steepest descent calculations where only the most probable
necklace conformation is accounted for are not always very accurate.

Let us briefly make a remark on single chain stretching in general. There came
up a number of experimental studies
\cite{lee:94.1,rief:97.1,senden:98.1,li:98.1,courvoisier:98.1}. Some of them
deal also with DNA stretching and defolding under forces
\cite{cluzel:96.1,bustamente:94.1}. Theoretical papers describe stretching of
DNA molecules \cite{marko:95.1}. The forces we are going to discuss in the
case of polyelectrolytes in poor solvent are much lower than appearing there.
In most cases related to the DNA overstretching plays a significant role, in
contrast to the problem described here.

The paper is organized as follows: in the next section we first summarize
briefly the known results on the pearl-necklace structure \cite{dobrynin:96};
we then discuss the structure of necklaces under an external force; we start
with the continuous limit, corresponding to many pearls connected by strings
and show that the pearls are "unwinded" by the external force; when the number
of pearls becomes small a more detailed treatment is necessary, we study in
details the dissolution of a necklace of two pearls. Section III is devoted to
a more general discussion of the discreteness of the number of pearls.  A
somewhat simpler description, that we think still captures the essential
physics, is developed to describe individual pearl unwinding for necklaces
involving many pearls.  Fluctuations progressively smear out force plateaus
associated to individual pearls when the number of pearls increases. This
whole analysis ignores the role of counterions of the chain. The pearl surface
potential however, can be rather high even for moderate charge fractions and
some charge regulation due to counterion condensation on the pearls is to be
expected. Large pearls inducing counterion condensation and very poor solvents
are discussed in section IV where we show that charge regulation can suppress
the large globule instability. We also consider very poor solvents for which
there is hardly any solvent in the pearl core and the dielectric constant is
low; the polyelectrolyte behaves then as an ionomer 
\footnote{A short account
  of this work has been presented at the St. Petersburg conference on
  ''Molecular Mobility and Order in Polymer Systems'' June 1999,
  \cite{conf:99}.}.

\section{Polyelectrolytes in poor solvent under external forces}
\subsection{Equilibrium necklaces, $\varphi = 0$}  Polyelectrolyte chains
in a poor solvent show a
rich conformational phase behavior because competing interactions are ruling
the chain statistics: surface
tension drives the globule toward small polymer/water interfaces  whilst
electrostatics
drive it toward large overall extensions. Therefore, the globular
structures for large enough
charge fractions are no longer
spherical as in the case of neutral polymers, but elongated. This was first
recognized by
Khokhlov \cite{khokhlov:80} who proposed a cylindrical globular shape
(assuming that there is only
one characteristic length). More precisely he finds that a globule of
radius $R$
becomes elongated when
$F_{\rm surface} = \gamma R^{2} > F_{\rm el} = \kt (\lb f^2N^2)/R$, i.e.,
$\tau > ({l_{\rm B}}/{b})f^2
N$. For larger charge fractions the globule elongates to a
thickness $D$ and length $L$.
This can be seen from a simple free energy balance. The free energy
of the cylindrical globule is
(remember that the
concentration is given by $c= \tau/b^3 = N/(D^2L)$) \begin{equation}
\label{cyl1}
\frac{F}{\kt} = \tau^{3/2} N^{1/2} \left( {L\over b}\right)^{1/2} +
\frac{\lb f^2 N}{L}
\end{equation} which yields upon minimization, the length of the cylinder
$L = b ({N / \tau})
\left( {\lb f^2}/{b}\right)^{2/3}$ and the diameter comparable to
the  size
$R_c\simeq b(\lb f^2/b)^{-1/3}$ of the
 marginally stable spherical globule, as already mentioned in the
introduction.

Kantor and Kardar \cite{kantor:95} have shown that the
elongated globule structure is unstable and breaks up into marginally stable
pearls by analogy with
the Rayleigh instability of charged liquid droplets. Due to the chain
connectivity
these globules are separated
by stretched strings. The details of the Rayleigh-like instability
for the
necklace structure
have been worked out by Dobrynin et al.
\cite{dobrynin:96} . The elongated
globule breaks up into pearls and strings. The pearls of size $R$,
contain $m = c_0 R^3$ monomers,
where $c_0 = \tau/b^3$ is the equilibrium concentration of the globule.
The elongated strings of length $l$ and
diameter
$d$, contain $m_{s}$ monomers. For completeness and
later use, we write down
the free energy in the following form
\begin{equation}
\label{1} {F\over \kt} = {N\over m} \left \{
\gamma R^2 + \frac{\lb f^2 m^2}{R} + \frac{F_{\rm elastic}(l)}{\kt} +
\gamma l d
+ \frac{\lb f^2 m_{s}^2}{l} \right\}
+\frac{lb f^2 N^2}{L}
\end{equation}
The terms correspond respectively to the surface energy of the pearls, to
the electrostatic self-energy of the pearls, to the elastic free energy of
the strings, to their
surface energy, to their
electrostatic energy and to the entire (smeared out) electrostatic energy
of the whole chain. Within
the limitations of this model, the necklace structure can be determined by
minimization of
eq.(\ref{1}). The size of the pearls is
$R \simeq R_{\rm c} = b/((\lb/b) f^2)^{1/3}$ and each pearl
contains $m = (\tau b)/
(\lb f^2)$ monomers. The
string diameter is $d = \xi_t$ and the string length $l= b (b\tau/ \lb
f^2)^{1/2}$. The total
length of the chain is $L= b N (\lb f^2/ b)^{1/2}
\tau^{-1/2}$. This last result
shows that the chain is indeed stretched. The physical picture can
be summarized as follows:
pearls of size $R_c$ are separated by strings with a tension equal
to the critical
tension for unwinding
pearls $\sigma$. This tension comes from the long range electrostatic
repulsions
between
pearls. The string length obtained from this argument
$l$ agrees with the minimization of the free energy. It is
easily checked that most
monomers are packed in pearls and that the stretched strings dominate
the total length $L$.

\subsection{Necklace stretching in the continuous limit, $\varphi \neq 0$}
\begin{figure}
\begin{center}
\begin{minipage}{12cm}
\centerline{{\epsfig{file=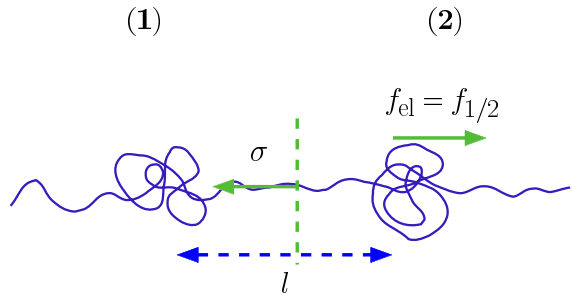,width=7.5cm}}} 
\vspace*{10pt}
\caption{\label{fig:necf2} 
  {\footnotesize The force balance on the half necklace {\bf (2)} : The
    tension {\bf t} balances the external force $\varphi$ and the
    electrostatic force exerted by half necklace {\bf 1}. Further assuming
    $t\sim \kt \tau/b$ and neglecting the electrostatic interaction between
    pearls far apart leads to the simple scaling argument for $l(\varphi)$ and
    $L(\varphi)$.  } }
\end{minipage}
\end{center}
\end{figure}

We first discuss qualitatively, the effect of an external force $\varphi$ to
the necklace chain.  When the polyelectrolyte necklace is stretched under an
external force, the pearl size and the string tension are essentially
unaffected as they are fixed by the Rayleigh instability and by the
equilibrium between pearl and string monomers respectively. The string length
$l$, is given by a force balance on a "half" necklace, relating the tension
$\sigma$, the external force $\varphi$ and the electrostatic force exerted by
the other half necklace, $\sigma =\kt \tau/b =\varphi + \kt \lb Q^2/l^2$; it
increases with the external force and thus some pearls are converted into
stretched strings (see Fig.\ref{fig:necf2}).

\begin{figure}
\begin{center}
\begin{minipage}{12cm}
\centerline{{\epsfig{file=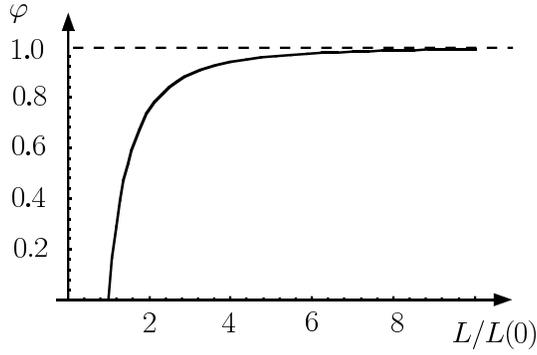,width=7cm}}} 
\vspace*{10pt}
\caption{\label{fig:necf3} 
  {\footnotesize The force law in the continuous model following the simple 
scaling argument. The length $L$ is reduced by the equilibrium length $L(0)$ 
and the plateau value of the force $\sim \kt \tau/b$ is taken as force unit. The end 
of this regime is reach when half of the pearls are unwinded, the elasticity is then dominated 
by the string elasticity and the force increases linearly (for gaussian elasticity),
this regime is not shown.  
} 
}
\end{minipage}
\end{center}
\end{figure}

This argument
yields  the pearl
separation length $l(\varphi)$ as a function of the external force. For
moderate external forces
where most monomers remain in pearls, we obtain the simple force law
\begin{equation}
\label{basic}
\frac{\varphi}{\kt} = \frac{\tau}{b}\left( 1 - \left( \frac{L(0)}{L}
\right)^2\right)
 \end{equation}
The corresponding force profile is represented in Fig.\ref{fig:necf3}.
This equation
holds for small enough values of $L$, i.e., $L < Nb \tau$ and
logarithmic factors have been
omitted. For larger forces, the hydrophobic forces are not important
and the elasticity is the Gaussian elasticity of the strings as for a
polyelectrolyte in a Theta solvent. The pearl number is treated here as
a continuous variable.
At small scales the force curve is expected to show plateaus
corresponding to the unwinding of individual
pearls, however for a large number of pearls, the curve is rounded by
fluctuations. The typical
fluctuation of the pearl number must then be larger than one.

To study the deformation behavior of isolated necklaces in more details we
use an explicit  free
energy close to  eq.(\ref{1}). A more detailed expression for the
interactions between strings
and pearls is introduced, and we further account for an imposed external
forces $\varphi$.
\begin{eqnarray}
{G \over \kt} &=& p \left(\frac{\tau}{b}\right)^2
R^{2} +  (p-1)
\left(\frac{\tau}{b}\right)^2 d l + p \frac{\lb f^2 m^2}{R} +
{1\over 2} \frac{\lb f^2
M^2}{L} \log(L/d) + \nonumber \\ &+& 2 \frac{\lb f^2 m M}{L}\log(L/R) +
p \frac{\lb f^2
m^2}{l} \log p + \frac{L^2}{2Mb^2} - \varphi L
\label{2}
\end{eqnarray}
 
We have used here the geometrical relationships for the
necklace structure,
the string length between two pearls is $l= L/(p -1)$, where $L$
is the total length of the
necklace chain and $p$ the number of pearls; the number of monomers
in one pearl is $m =
(N-M)/p$, where $N$ is
the total number of monomers in the entire chain and $M=(p
-1)m_{s}$ is the total number of monomers in the strings; the diameter of the
strings is given by $d^2 =
Mb^3/L \tau$; finally the radius of the pearls is $R = ((N-M)/p)^{1/3}
b/\tau^{1/3}$. It is useful to note that despite we have written 
out the logarithmic corrections here, we do not derive with respect to
the chain length them in the 
following. We added them to keep control
on the physical origin different terms. Therefore
we treat them as constants. Precise numerical factors, however, are omitted
as well.

It turns out to be convenient to use again the  parameter $\Lambda=b\tau^3/\lb
f^2$ proportional to the number of thermal blobs in
a pearl and
 the following dimensionless
variables, a reduced tension $t = L/(Mb\tau)$ and
a reduced external
force,  $\tilde \varphi = \varphi b/\tau$. This amounts to rescale
every quantity
by the thermal blob. The physically relevant cases
correspond to $\Lambda >>1$,
 thus $1/\Lambda$
can be used as an expansion parameter. With these definitions we rewrite
the free energy as

\begin{eqnarray}
\label{21}
\frac{G}{\kt} &=& p^{1/3}(\tilde N - \tilde M)^{2/3} +  {1\over
\Lambda p^{2/3}}(\tilde N
- \tilde M)^{5/3}+ \nonumber \\  &+& {1\over {2 \Lambda t} } \tilde M \log
\left({L/d}\right) + 2 {1\over \Lambda
t} (\tilde N - \tilde M) \log(L/R) + \nonumber \\ &+& {1\over \Lambda t} \frac{(\tilde
N - \tilde M)^2}{\tilde
M}\log p \nonumber \\ &+& \tilde M t^{1/2} + \tilde M t^2 - \tilde \varphi
\tilde M t
\end{eqnarray}
The free energy is expressed only in terms of
scaled variables and we
introduced $(\tilde N,\tilde M) = (N,M)\tau^2$. The independent variables
for the
minimization are
$\{p, \tilde M, t \}$. After minimization of the free energy, we obtain
\begin{equation}
\frac{\tilde N - \tilde M}{p} = {\Lambda \over 2}  
\end{equation}
\begin{eqnarray} 
-{2\over 3}
\left( \di \right)^{-1/3}
-{5 \over 3\Lambda} \left( \di \right)^{2/3} - \left( \frac{\tilde N -
\tilde M}{\tilde M} \right)^2
\frac{1}{\Lambda t} \log p -2 \frac{\tilde N - \tilde M}{\tilde M}
\frac{1}{\Lambda t} \log p \nonumber \\  
+ \sqrt{t} + {1\over 2} {1 \over {\Lambda t}} \log
(L/d) - {2\over \Lambda t}
\log (L/R) + t^2 - \tilde \varphi t = 0 \\
 2t + \frac{1}{2 \sqrt t} - \frac{1}{\Lambda t^2} \left(  2 \left(
\frac{\tilde N - \tilde
M}{\tilde M} \right) \log(L/R) + \left( \frac{\tilde N - \tilde M}{\tilde
M} \right)^2 \log p \right) - \tilde \varphi - \frac{1}{2 \Lambda t^2} \log
(L/d) = 0
\end{eqnarray}
 The first equation
determines the number of blobs in one pearl, in agreement with our previous
arguments. The second derivative with
respect to $p$ allows to evaluate the fluctuation of the number of
pearls, we find a r.m.s fluctuation
$\delta p= \Lambda^{1/3}/p^{1/2}$ (we neglect here the  coupling
between the fluctuations of $p$
and the fluctuations of other variables but
 this does not qualitatively change the result). The continuous
description is consistent for
$p>\Lambda^{2/3}$ when the fluctuation of the pearl number is larger than
one.
The two other equations determine the effective tension $t$ and the
number of monomers in the strings or equivalently the ratio $\Delta$ of the
number of monomers in the pearls to that in the strings $\Delta = (\tilde N
- \tilde M) /
\tilde M$.
 \begin{eqnarray}
\label{forcedelta}
 -{3\over 2} \left( \frac{2}{\Lambda} \right)^{1/3} - \frac{1}{\Lambda t}
\Delta (\Delta +2)\log
p + \sqrt t + {1\over {2 \Lambda t}} \log(L/d) + \frac{2}{\Lambda
t} \log(L/R) + t^2 - \tilde
\varphi t = 0\\  - {1\over \Lambda t^2} \Delta \left\{ \Delta \log p + 2
\log (L/R)
\right\} + 2t + \frac{1}{2 \sqrt t} - \tilde \varphi - \frac{1}{2 \Lambda
t^2} \log(L/d) = 0
\end{eqnarray}
The external force can be eliminated and
\begin{equation}
\label{tension}
\frac{1}{\Lambda t}\log(L/d) + \frac{2 \Delta}{\Lambda t} \log \left(
\frac{L}{Rp} \right) + \frac{2}{\Lambda t} \log(L/R)
-t^2 + \frac {\sqrt t}{2} - {3\over 2} \left( \frac{2}{\Lambda}
\right)^{1/3} = 0
\end{equation}
For the free necklace , $\varphi = 0$, the fraction $\Delta$ of pearl
monomers to string monomers
is of order $\Delta\sim \sqrt{\Lambda/\log p}$. When a  finite
force is applied, $\Delta$ decreases.
In the limit $\Lambda \to \infty$, the reduced tension is $t = t_0 =
(1/2)^{2/3}$, which can be used as a  basis for an expansion to find
approximate solutions.
Following eq. (\ref{tension}), the leading correction to $t_0$
comes from the the pearl self-energy; it is negative
and proportional to
$\Lambda^{-1/3}$. It is a typical finite pearl size correction proportional
to $\xi_t/R$ and
comprising both surface tension and electrostatic contributions of the same
order.
The next order correction comes from the electrostatic interactions, it is
proportional to
$\Delta$ and positive;
 at small force, it is proportional to $\Lambda^{-1/2}\log(L/R)/\log{p}^{1/2}$.
 At moderate force,  the force extension curve to leading order in
 $1/\Lambda$, is given by
\begin{equation}
\label{forcepearl}
\frac{\varphi}{\kt} = (2t_0+1/2\sqrt {t_0}){\tau\over b}\left(1- \alpha
\Lambda^{-1/3}\right) - \frac{\lb f^{2}N^{2}}{L^{2}}
\log p
\end{equation}
 where $\alpha$ is a positive numerical constant, depending on $t_0$ only.
This last equation
corresponds now to eq.(\ref{basic}).
 Physically it describes the unwinding of pearls by increasing external
forces.
Two new features appear in comparison with the simple argument above. The
first is that a
new term which corrects the (bare) line tension $\sigma = \kt \tau/b$  by a
term
proportional to $\Lambda^{-1/3}$. This term has its physical origin in the
finite size of the pearls, it is
 proportional to $(\xi_t/R)^{1/3}$ as for  the unwinding of a neutral
globule but has both surface tension
and electrostatic contributions of the same order.
 The other
point is the appearance of the logarithmic term, which comes from the
interactions between pearls
that are dominant. Under the logarithm, the number of pearls can be
approximated by $p =
(N\tau^2 - L\tau/b)/\Lambda$.

When a finite fraction of the pearls are unwinded $\Delta$ becomes of order
unity. The
force law eq.(\ref{forcepearl}) can no longer be applied when $ L/\tau N
b\sim 1$.
 The force is  then a second order
polynomial in $\Delta$ and,
due to the correction to $t_0$ (see eq.{\ref{tension}),
the linear term in $\Delta$ is positive; $\varphi$ is of the form
$\varphi = A - B\Delta^2 + C\Delta$ where all three constants are positive
and $A$ and $B$ are of the same
order. This  leads to a spinodal instability when about half of the
pearls are unwinded.
We thus expect the polymer to jump from a metastable necklace to a single
string unwinding about
half of the pearls at once.  Note that our argument relies on small
correction terms to the force
and tension (as $\Delta\sim 1$); our free  energy, involving several
approximations,
may be too crude in this case. However when half of the pearls are unwinded
the elasticity is
dominated by the string elasticity even in the absence of
any
transition.

\subsection{Discrete pearls: two pearl-necklace}

When there are only a few pearls in the necklace, the fact that the
number of pearls is discrete cannot be ignored. This is of particular
importance if the total number of pearls is small as shown by many
numerical simulations \cite{stoll:99,kremer:99,duenweg:99}.
In this subsection, we discuss the special case of two pearls
separated by a string.
This is the simplest model which is physically meaningful. For energetic
reasons based on the here used free energy, 
the ends of the necklace chain consist  of pearls (a chain with one
pearl in the middle and
two strings at the end is energetically not favorable). This minimal model
has the advantage that
the dissolution of the pearls can be computed in details. The number of
pearls is fixed, in
contrast to the continuous model.
If the necklace
had more than two pearls, we would have to take into account the fact
that the pearls at the chain end,  are larger than the central pearls,
simply because
of smaller
electrostatic interactions
(one missing neighbor). For a small number of pearls, we expect a
discontinuous stretching with force
plateaus corresponding to the unwinding of a pearl.

In order to describe the jump from 2 to 0 pearls, we start from a free
energy that takes
into account the
interactions more precisely and includes
all the surface and electrostatic free energies.
\begin{eqnarray} G= 2\gamma R^{2} &+& 2 \frac{\lb f^{2}m^{2}}{R} + \gamma Ld +
\frac{\lb f^{2}M}{L}\log(L/d) + \nonumber \\ &+& 2 \frac{\lb
f^{2}mM}{L}\log(L/R) + \frac{\lb
f^{2} m^{2}}{L}+\\ &+& \frac{L^2}{2Mb} - \varphi L \nonumber \end{eqnarray}
The free energy can be
rescaled using dimensionless variables ($\tilde m= \tau^{2}m$):
\begin{eqnarray}
\label{twopearl}
\frac{G}{\kt \tilde N}& =& {1\over \tilde N} \left( 2 \tilde m^{2/3}+
{2\over \Lambda} \tilde
m^{5/3}\right) + {1\over \Lambda t} {2 \tilde m\over \tilde N} \log(L/R) +
\nonumber \\ &+ &{1\over
\Lambda t}
\left(1-\rr \right) \log(L/d) + \left(1-\rr \right) t^{1/2} + \left(1-\rr
\right) t^{2}+\\ &+&
{1\over \Lambda t} \left( {\tilde m\over\tilde N} \right)^{2}
\frac{1}{\left(1-\rr \right)} - \tilde\varphi t
\left(1-\rr \right) \nonumber \end{eqnarray} As the number of pearls is
fixed, there are
only two  variables: the number of thermal blobs in a pearl $\tilde
m$ and the reduced tension $t$. Two simple limits are described by
this equation:
 $\tilde m \to 0$ corresponds to a string with no pearls and the chain is
 an extended string of thermal blobs in the so-called
Pincus regime \cite{degennes:79}; in the absence of external
force, $\varphi = 0$, we describe a free two-pearl necklace.

The free energy map $\mbox{G}(t,m)$ eq.(\ref{twopearl}) is shown in
Fig.\ref{fig:necf4}, for $\tilde N = 2.05 \Lambda/2$ with $\Lambda = 10^3$.
The plot is typical for high $\Lambda$-values where the description by the
most likely necklace (steepest descent) is meaningful.
\begin{figure}
\begin{center}
\begin{minipage}{12cm}
\centerline{{\epsfig{file=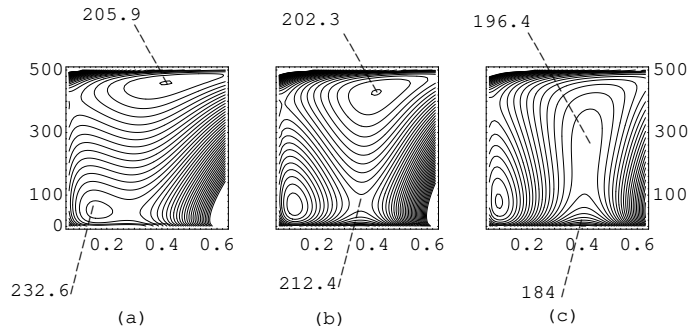,width=9cm}}} 
\vspace*{10pt}
\caption{\label{fig:necf4} 
  {\footnotesize 
G-map  in the $t, m$ plan for increasing applied forces.
The preferred pearl size is $500$. The Khokhlov local minimum (low $m$, low $t$),
 is not shown. 
 {\bf a} In addition to the necklace minimum,
an open string minimum appears (this minimum becomes deeper for higher applied forces)
 ; {\bf b} The applied force is such that the two minima have the same depth. They are separated 
by a saddle point about $10\kt$ higher in energy.{\bf c} The metastable necklace 
minimum disappears.        
} 
}
\end{minipage}
\end{center}
\end{figure}
At low force, there is only
one minimum in the $G$-map that corresponds to a necklace with roughly the
optimal pearl
 size. For higher
applied force the pearls unwind. At  some value of the force, situation
(a), a second minimum corresponding
to a stretched  string without pearls appears as a metastable state. With
increasing force
the free energy $G$ of the stretched string
 becomes closer to the necklace free  energy, the two minima in the
 free energy are equal
in situation (b); they
are separated by a saddle point significantly higher in energy. For even
higher forces the necklace becomes
 metastable; the necklace minimum disappears in situation (b) where
 roughly
half of the pearl content
is unwinded. From these simple considerations, we  plot an elongation/force
diagram in Fig.\ref{fig:necf5}with
two branches. The stability exchange (situation (b) in the contour plots)
is shown by the dashed line.
\begin{figure}
\begin{center}
\begin{minipage}{12cm}
\centerline{{\epsfig{file=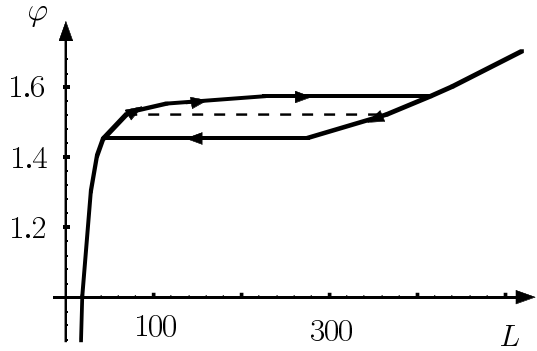,width=7cm}}} 
\vspace*{10pt}
\caption{\label{fig:necf5} 
  {\footnotesize 
The two pearl necklace branch and the open string branch of 
the length/force diagram. The values shown correspond to the minima in the previous 
G-map (Fig.\ref{fig:necf4}). The dashed line correspond to the stability exchange (b). The largest 
hysteresis loop  is shown.        
} 
}
\end{minipage}
\end{center}
\end{figure}
Below the dashed line the stretched string branch is metastable whereas
above this line, the necklace branch is metastable. In
a fast stretch/collapse/stretch cycle we expect a hysteresis loop following
most of the metastable branches. 
In a quasi static extension procedure under imposed
increasing force  a pseudo-plateau close to
the dashed line in the force-extension curve is expected. A better
description of the pseudo-plateau requires a
more complete theory including thermal fluctuations. A first
approximation for the chain partition
function is:
\begin{equation}
\Xi = \int dm\,dL \exp (-G(m,t)/\kt )
\end{equation}
where the integral is carried out over the previous ${t,m}$ map. The
average length is obtained as the derivative of the logarithm of the partition
function $\langle L \rangle =\frac{ \partial \log \Xi}{\partial {\varphi}}$.
The force-extension curve is plotted in Fig.\ref{fig:necf6}. 
\begin{figure}
\begin{center}
\begin{minipage}{12cm}
\centerline{{\epsfig{file=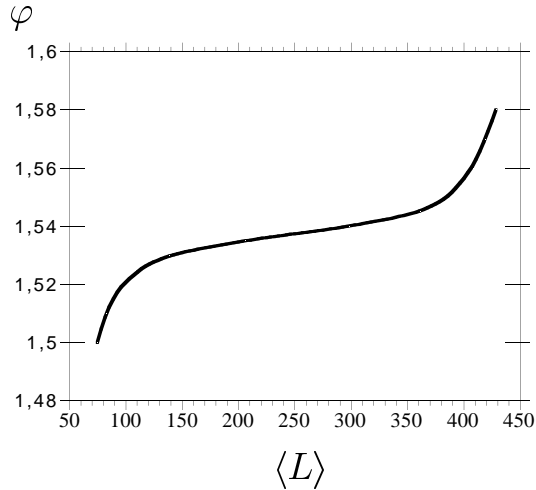,width=7cm}}} 
\vspace*{10pt}
\caption{\label{fig:necf6} 
  {\footnotesize 
Due to fluctuations a force pseudo plateau (rather than a plateau) 
is found. We plot the force $\varphi$ against the corresponding average 
length $\left< L \right>$. The force at the pseudo plateau (say inflection point) is close to the 
stability exchange value reported in Fig.\ref{fig:necf5}, the pseudo plateau extends roughly 
between the same length as the dashed line in Fig.\ref{fig:necf5}. 
In the pseudo plateau, the length 
fluctuations are comparable to the average length.    
} 
}
\end{minipage}
\end{center}
\end{figure}

A pseudo-plateau is
indeed observed. According to linear response theory, the length
fluctuations are obtained from the slope of this curve
$\langle L^2 \rangle - \langle L \rangle^2=
kT \frac{\partial  \langle L \rangle}{\partial \varphi}$. The RMS fluctuation
of the length is of
the order of the average length on the pseudo-plateau and much weaker
outside the
pseudo-plateau.
This description remains crude as we did not allow for unlike pearl
sizes. As shown
in the simple discussion of the Rayleigh instability for mesoscopic
drops, pearl size fluctuations may be
important for not too high values of  $\Lambda$.
In the following, we give a simpler description where the tension of
the string is kept constant
 during the pearl unwinding (a reasonable assumption in view of the
numerics). It is then possible to
get a simple picture also in the case when the necklace has several pearls.

\section{Discrete pearl model}

So far,  we have ignored pearl size fluctuations; in the following we propose
a simplified description of the necklace structure that allows to
take these fluctuations into account.
We describe approximately the necklace by a row
of pearls, and we fix the structure of the strings  i.e. their tension
and their elastic energy per unit length. If the total number of monomers in
the strings is $M$ as above and the total length of the chain (dominated by the
strings) is $L$, $M/L= \tau/b$ and the total elastic energy of the strings is
$F_{\rm string}\sim \frac1 2 \kt L\tau/b$.

 We then distribute the $P=N-M$ remaining monomers into $p$ pearls with masses
$m_{i}$, which are not necessarily identical. The total free
energy includes the pearl self-energy, the string electrostatic
energy and the smeared out
electrostatic energy $\kt \lb f^2N^2/L$ between pearls and strings.
In this model, all the contributions to the free energy except for the
pearl self-energy are identical to those of the continuous
model. We do not include here the logarithmic factors in the
electrostatic term but they could be included as before. For a fixed
length of the chain, the number of monomers in each pearl and the
fluctuations of this number are essentially determined by the pearl
self-energy, we thus ignore the fluctuations in all the other terms of
the free energy. The main approximation of our model is that the
length of the chain is only due to the strings and thus that the total
number of monomers in the strings and therefore in the pearls are fixed.
This approximation may not be quantitative but we expect it to give
the correct physical picture and the correct scaling behavior.

In this section, in contrast to previous considerations, we impose the length
of the chain and
compute the force.
In order to discuss this model qualitatively, we further simplify it
by assuming that all pearls are identical.
In a second step the fluctuations in the number of monomers of the
individual pearls are explicitly taken into account.

The free energy of a single pearl is
\begin{equation}
\label{pearlss} 
{F_{p}} = \gamma R^2 + \kt \frac{\lb f^2 m^2}{R} =
\kt \left( N^{2/3} \tau^{4/3} + {\lb\over b} f^2 \tau^{1/3} m^{5/3}\right)
\end{equation}
 where $m$ is the number of monomers in the pearl. The free energy per pearl monomer is then
 \begin{equation} {\tilde f} = m^{-1/3}
\tau^{4/3} + {\lb \over
b}f^2 \tau^{1/3} m^{2/3} \end{equation}
The free energy per monomer
${\tilde f}$ plotted as a
function of the pearl mass $m$ has a minimum at $m_0 = \Lambda /(2
\tau^2)$. The second derivative is positive for $m=m_{0}$, it vanishes
for $m= 4m_0$. In the
following, in order to take into account the deviations from the
preferred value $m=m_{0}$, we approximate the free energy per monomer by a
parabolic form where
the parabola has a spring
constant $k=(\partial^2 {\tilde f}/\partial m^2) \vert_{m_0} =
m_0^{-1}(\lb f^{2}/b)^{4/3}$. The total free energy of the chain can
then be written as
\begin{equation}
F= F_{\rm cont}(m_{0})+ \frac 1 2 \kt km_{0}\sum_{i=1}^{p} (m_i-m_{0})^{2}.
\end{equation}
where $F_{\rm cont}$ is the free energy discussed earlier in the
continuous limit with equal numbers of monomers $m_{0}$ in all
pearls $(F_{\rm cont}$ is in a first approximation given by eq. (\ref{1})
or by eq. (\ref{2}) where the force term has been omitted)
and
the sum is over the $p$ pearls.

As a first step, we consider that all the pearls have the same number of
monomers
$m$ that can be different from the preferred value $m_{0}$. When the
necklace is stretched, the number of monomers in each pearl
decreases. At some point however, it becomes
energetically more favorable to have one pearl less. In this crude argument,
we assume that the jump
from $p$ to $p-1$ pearls occurs abruptly for a given imposed necklace
length, this is equivalent
to a large energy scale in the Boltzmann weights and hence to large pearls
($\Lambda>>1$). The stability exchange
from the necklace containing $p$ pearls to the necklace containing $p-1$
pearls corresponds, in the parabolic potential, to opposite
deviations from the preferred pearl size: $\delta m_{p} = -\delta m_{p-1}$.
On the other hand, due to mass
conservation, (the length is imposed), $p m_{p} = (p-1) m_{p-1}$.
The necklace thus jumps from $p$ to $p-1$ pearls for   $\delta m_{p} = -
m_0/(2p-1)$.
 For a chain containing $p$
pearls, we get
$$
\delta m = {1\over p} \left( N - {L\over {b\tau}} - pm_0\right) $$
The force is obtained by derivation of the total free energy, it is
equal to
\begin{equation}
\varphi_p =\varphi_{\rm cont} - {1\over p} \frac{km_0}{\tau b} \left( N -
{L\over {b\tau}} -
pm_0\right)
\label{steepforce}.
\end{equation}
The first term is the force calculated previously in the continuous limit and
the second term is the contribution due to the discreteness of the pearls and
to the fact that they do not have the preferred size; this "individual pearl"
contribution is non monotonic (see Fig.\ref{fig:necf7}).  If the number of
pearls in the necklace is small enough, the total force also exhibits
downwards jumps.
\begin{figure}
\begin{center}
\begin{minipage}{12cm}
\centerline{{\epsfig{file=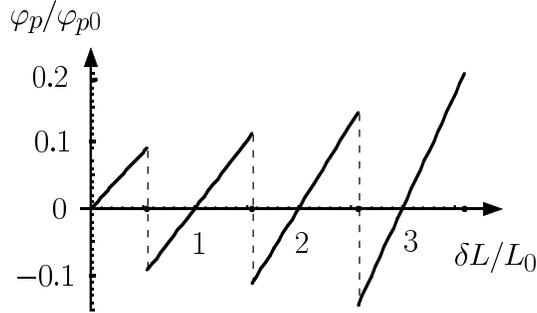,width=7cm}}} 
\vspace*{10pt}
\caption{\label{fig:necf7} 
  {\footnotesize 
The discreteness of the pearl number introduces a non 
monotonous force correction on top of the continuous force. Here this correction 
is evaluated in the simplified approach neglecting fluctuations. A unwinded pearl
 contributes a length $l_0$.The origin on the x-axis corresponds to a happy 
six pearl necklace.  The force unit is chosen as $\kt\tau/b \Lambda^{-1/3}$   
} 
}
\end{minipage}
\end{center}
\end{figure}
This occurs if the distance $l$ between pearls is
large enough ($l>l_0\Lambda^{-2/9} m_0\tau b$).
 The branches where the force decreases with length are unstable and
the force actually shows a plateau. The plateau is obtained from a
Maxwell-like construction on the total
force.
In practice, for non quasi-static transformations, the system  rather
follows the metastable branch
 and, when the applied force increases, the length jumps at the force
extrema.

The force jumps obtained  upon imposing the chain length are rounded by
fluctuations of both the number of monomers in a pearl and the number of pearls.
In order to take into account the fluctuations, we compute the
partition function of a necklace of $p$
pearls
\begin{eqnarray} Z_p = Z_0 \int \prod_i dm_i \exp
\left\{ -{1\over 2} k m_0
\left(m_i-m_0 \right)^2 \right\}
\delta \left( N - {L\over {\tau b}} - \sum_{i=1}^p m_i \right).
\end{eqnarray} The delta
function ensures the mass conservation. The bare partition function is
associated to the free energy calculated in the continuous  limit
$Z_{0}=\exp (-F_{\rm cont}/\kt)$.
Formally we are looking for the partition function of a row
of $p$ harmonic springs
with a fixed total length. The
result is
\begin{equation} Z=Z_0 \left( \frac{2\pi}{km_0} \right)^{p/2} \left(
\frac{2\pi km_0}{p}
\right)^{1/2} \exp \left\{ -\frac{km_0}{2p} \left(N -{L\over{\tau b}} -
pm_0 \right)^2 \right\}
\end{equation} The total partition function of the necklace chain is
$Z=\sum_{p} Z_{p}$ and the free energy is $F=-\kt \log Z$.
The force is then obtained by derivation
\begin{eqnarray}
\label{huhu}
\varphi = \varphi_{con} +
\kt\frac {\sum_p \left( \varphi_p {(\lb f^{2}/b)^{-2p/3}\over p^{1/2}}
\exp\left(-{\Lambda^{2/3}\over p} (N/m_0
-L/(\tau bm_0) -p) \right)^2 \right)} {\sum_p \left( {(\lb
f^{2}/b)^{-2p/3}\over p^{1/2}}
\exp\left(-{\Lambda^{2/3}\over p} (N/m_0 -L/(\tau bm_0) -p) \right)^2
\right)}, \end{eqnarray}
where
$ \varphi_{con}$ is the force obtained in the continuous
approximation \ref{forcepearl} and
\begin{equation}
\varphi_p = -2\frac{\tau }{b } \frac{\Lambda^{-1/3}}{p}(N/m_0-L/\tau b
m_0-p)
\end{equation}
The fluctuations of the pearl number smear out the force oscillations for
$p>\Lambda^{2/3}$ as found previously. The fluctuations of the pearl sizes are
reflected by the prefactors $p^{-1/2}(\lb f^{2}/b)^{-2p/3}$.  The sums are
evaluated numerically for certain values. The result is displayed on
Fig.\ref{fig:necf8}. As in our simple argument where the fluctuations were
ignored, the force shows oscillations with unstable decreasing branches that
should be replaced by plateaus or pseudo-plateaus using the Maxwell
construction.  In the limit of very large $\Lambda$, the decrease of the force
becomes very steep and the force on each branch tends to the value given by
equation \ref{steepforce}.
\begin{figure}
\begin{center}
\begin{minipage}{12cm}
\centerline{{\epsfig{file=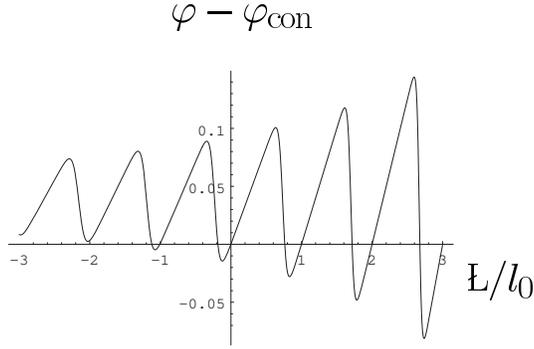,width=7cm}}} 
\vspace*{10pt}
\caption{\label{fig:necf8} 
  {\footnotesize 
Same as Fig. \ref{fig:necf7} but with fluctuations taken into 
account. We take $\Lambda = 1000$ and $\lb f^2/b = 10^{-4}$. 
For such high $\Lambda$-values fluctuations have 
not a dramatic effect when there are only a few pearls, 
for $\Lambda = 100$ (not shown)
the fluctuations almost suppress the first oscillations completely  
 (left of the figure).   
} 
}
\end{minipage}
\end{center}
\end{figure}

\section{Localized counterions: very poor solvent and counterion
  condensation on the pearls}

So far, the counterions of the necklace chain have not been taken into
account and have been considered as forming a free ideal solution.
There are however  cases where this assumption fails.

When the solvent is very poor, the pearls are very dense and the local
dielectric constant inside a pearl is lower than that of water by typically
one order of magnitude, there are essentially no dissociated charges in the
pearl core.  Charged systems with low dielectric constants, called ionomers,
have been extensively studied \cite{ionomers}.  We follow in this section the
lines of a simple model proposed by Dobrynin \cite{dobrynin:99}.

The electrostatic potential at the surface of a pearl $\sim \tau
(\lb/f)^{1/3}$ increases with decreasing solvent quality and decreasing charge
fraction. A condensation of the counterions and thus a regulation of the
charge of the pearls is expected when this potential becomes of order $\kt$
i.e.  when $\tau^3>f b/\lb $.  The necklace becomes then unstable and
collapses to form a compact globule.  We now discuss this instability and the
effect of an external force.

\subsection{Condensation on the pearls}

We assume that the counterions condense to a certain amount on the
pearls but that they do not condense on the strings,
i.e., the linear charge density along the strings is smaller than the
Bjerrum length, which leads to
$f<\tau (b/\lb)$. Counterion condensation then only takes place
on the pearls.
In the simplest description \cite{oosawa:71}, the pearls have an effective charge
$Q_{\rm e}$ determined by equilibrating the chemical potential of the
counterions to bulk value of the order of the thermal energy $\kt$.

The pearl-necklace structure is based on  a Rayleigh instability. We first
discuss the effect of condensation for a "liquid of disconnected blobs".  Let
us call $-\alpha \kt$ the chemical potential of the free counterions and $\qe$
the net charge of a droplet of radius $R$ and of nominal charge $Q = f\tau
R^3/b^3$, the free energy of an assembly of $\tilde N$ blobs forming p
monodisperse non-interacting droplets of radius $R$ is $$F= p \gamma R^{2}+\kt
p \frac{Q_{\rm e}^{2}\lb}{R}+ p \alpha \kt (Q-Q_{\rm e}).$$
It includes the
surface tension energy, the electrostatic energy of each drop and a chemical
potential term for the counterions. After minimization with respect to the
effective charge and taking into account the fact that the effective charge
must be smaller than the nominal charge we obtain
\begin{eqnarray}
\label{RaylCond}
F\tau^3/\tilde N\kt &=& {\tau^2 b\over R} + {l_B f^2 R^2\tau^2\over 2
b^3}\quad (R<R_{co})\nonumber\\
F\tau^3/\tilde N\kt &=&  {\tau^2 b\over R} - {\alpha^2 b^3\over 2 l_B R^2}
+ f\tau\alpha
\quad (R>R_{co})
\end{eqnarray}
charge regulation takes place for $R>R_{\rm co}$ where 
$R_{\rm co}^2 = {(\alpha b^3) / (f\tau \lb)}$.

If the fraction of charged monomers is large enough, $\alpha^3bf/l_b\tau^3>1$,
in the regime where there is no counterion condensation,
the free energy per blob
 has a minimum
at $R_{min} =
b(l_Bf^2/b)^{-1/3}$ (the electrostatic blob size). The associated droplet
radius is such that
$R_{min}<R_{co}$ and the minimum actually exists. In the same range
of parameters, in the regime where the counterions condense on the spheres
the free energy per blob
 has a maximum in the
charge regulated regime at $R_{\rm max} = \alpha^2b^2/\tau^2\lb$. The minimum
at the preferred pearl size is then
stable against the infinite globule.
In the opposite limit $\alpha^3bf/\lb\tau^3<1$, the free energy per blob
decreases monotonically with the radius of the droplets and the
largest possible globule is stable. In this case, counterion
condensation prevents the Rayleigh instability. The
energy per blob is sketched in
Fig.\ref{fig:necf9} for both cases.
\begin{figure}
\begin{center}
\begin{minipage}{12cm}
\centerline{{\epsfig{file=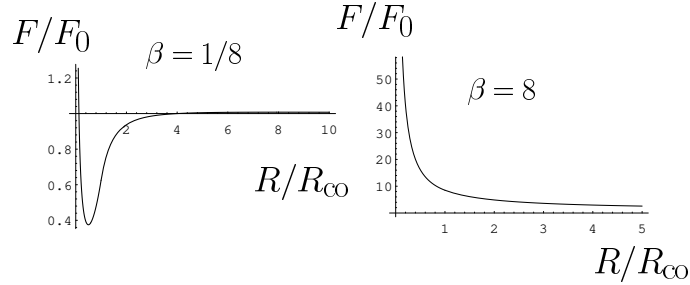,width=9cm}}} 
\vspace*{10pt}
\caption{\label{fig:necf9} 
  {\footnotesize Charge regulation by the pearls. Free energy per blob as the
    function of droplet radius for an assembly of non interacting droplets,
    the infinite globule is taken as reference state. The radius is reduced by
    $R_{\rm co}$, the crossover radius to charge regulation. The parameter
    $\beta = \left((\lb \tau^3) / (b \alpha^3 f )\right)^{1/2}$ controls
    regulation by the preferred pearl.  The plotted function reads:
    $f(x,\beta) = \beta/x +x^2/2 \quad (x<1)$ and $f(x,\beta) = \beta/x -x^2/2
    +1 \quad (x>1)$ For $\beta<1$ there is a preferred pearl located in the
    non regulated regime, the maximum located in the regulated regime is very
    shallow (the figure corresponds to $\beta = 1/8$).  For $\beta>1$, there
    is no finite preferred pearl size, the largest possible globule is stable
    (the figure corresponds to $\beta = 8$).  } }
\end{minipage}
\end{center}
\end{figure}

The structure of a necklace when there is no counterion condensation on the
pearls has been described in the previous sections. The argument on the
Rayleigh instability suggests that charge regulation and collapse occur
simultaneously, as stated by Dobrynin et al.\cite{dobrynin:99} and Schiessel
and Pincus \cite{schiessel:98} following similar arguments.  The situation is
also reminiscent of that of cylindrical globules studied by Khokhlov
\cite{khokhlov:80} where Manning condensation \cite{manning:} provokes
collapse.

We now turn to the case of a polyelectrolyte chain and consider a necklace
structure with condensed counterions on the pearls.  The total free energy can
be written as
\begin{eqnarray}
\label{fqe} {G\over \kt} = p\gamma R^2 + (p-1) \gamma ld +
p \frac{\lb
\qe^2}{R} + \nonumber \\
 +(p-1) \frac{\lb f^2 m_s^2}{l} &+& \frac{\lb p^2
\qe^2}{L}\log p + \frac{L^2}{Mb^2} +
\alpha p(Q-\qe) -\varphi L
\end{eqnarray} where the first term is
the surface contribution of the $p$ pearls, the second term the surface free
energy of $p-1$ strings of length $l$ and thickness $d$, the third term is the
electrostatic self-energy of the pearls, followed by the electrostatic
self-energy of the strings containing $m_s$ monomers. The fifth term is the
total electrostatic interaction between the pearls where the total length of
the necklace is $L \simeq (p-1) l$.  The total elastic free energy
contribution yields the Gaussian term and finally the next term is the
contribution of the free counterions, the last term is the contribution of the
external force.  The minimization of the free energy eq.(\ref{fqe}) with
respect to $\qe$ gives the effective charge as a function of the radius $R$ of
the pearls and of the distance $l$ between pearls , the onset of charge
regulation defined as $\qe = Q$ is located at $Q = {\alpha R\over
  l_B(1+\frac{2R}{l+R}\log p)}$.  Due to the interactions between pearls, the
condensation threshold is shifted to a lower pearl charge; the effect depends
on the force through $l$ and $p$ and decreases for increasing forces. However
this is a small correction for high values of $\Lambda$. Ignoring these
corrections, the free energy can be minimized using the same set of variables
as in section (II-B).  The free energy does not have a minimum at any finite
value of $p$ and is minimum for $p=1$ or $p=0$. As for the disconnected liquid
of blobs, the necklace structure is unstable and as soon as the counterions
condense on the pearls, the polyelectrolyte collapses.  Within this
description there is no regime of necklace stabilized by an external force.
We thus conclude that at low charge fraction $f<\tau^3$ where there would be
charge regulation by the pearls, the dense globule is stable and stretches
discontinuously when the force is equal to $\kt \tau/b$, the necklace
structure is never stable.

\subsection{Very poor solvent}

In the preceding sections, we have assumed that the solvent is poor, but that
the reduced excluded volume $\tau$ is smaller than unity, i.e., the
correlation length remains much larger than the monomer size.  In this case,
one can assume that locally the dielectric constant is that of water, that the
Bjerrum length has the water value and that the charges are dissociated.
Consequently the counterions can be distributed inside the globule, almost
freely. In a poorer solvent, the correlation length $\xi_{t}$ is of the order
of the monomer size $b$ and this assumption on the dielectric constant cannot
be made. Inside the dense phase, the solution is close to a polymer melt and
the Bjerrum length decreases by one order of magnitude.  This is also true
well inside the pearls; the charges are then not dissociated and a typical
ionomer behavior is expected. In the following brief discussion we assume that
ion dissociation takes place only at the surface of the globules, and that
only those groups localized in a thin shell of the order of the thermal blob
size $\xi_{t}$ are dissociated and contribute to the pearl net charge.

The effective charge therefore becomes
\begin{equation}
\qe = fR^{2}\xi_{t} c = f \frac{R^{2}}{b^{2}},
\end{equation}
 where the concentration inside the globule is $c\propto b^{-3}$.
 The balance between the electrostatic
free energy due to this
effective charge and the surface free energy yields the size of the pearls
\begin{equation}
 R = b {b\over \lb} \left({\tau \over f}\right)^{2}.
\end{equation}
which enables us to compute the effective pearl charge and pearl electrostatic
potential:
\begin{eqnarray}
\qe = \left( {b \over \lb}\right)^{2} \tau^{4} f^{-3}, \\
 U_{\rm s} = (\lb
\qe)/ R \sim \frac{\tau^{2}}{f}
\end{eqnarray}

These expressions are only valid in the range of values of $\tau $ close to 1
and if part of the dissociated ions do not condense in the outer pearl shell,
i.e., $f>\tau^{2}$. On the other hand, the pearl size $\xi_{t}$ must be larger
than the string, diameter i.e., the thermal blob size $\xi_{t}$, which yields
an upper bound of the form $f< \tau^{3/2} (b/\lb)^{1/2}$. The charge fraction
must thus be in the range
\begin{equation}
\tau^{2} < f < \tau^{3/2} (b/\lb)^{1/2}. \end{equation}
The number of
monomers in the pearl is then
$m = R^{3}c$, giving $$ m = \left( {b\over
\lb}\right)^{3}
\frac{\tau^{7}}{f^{6}}. $$ Again the string length between two pearls can
be computed by the force
balance. It is given by
$$
l = b \left( {\lb \over b}\right)^{-3/2} \left({\tau^{7}\over
    f^{6}}\right)^{1/2} 
$$
and the calculation of the mass of the strings
$m_{\rm s} = l d^{2}c$ shows that still most charges are indeed in the pearls.
If an external force is applied, the force balance has to be modified and the
force-extension law is
\begin{equation} L = N \left( \frac{b}{\tau / b - \varphi} \right)^{1/2}
\left( {\lb \over
b}\right)^{3/2} {f^{6}\over \tau^{6}}. \end{equation}
This last equation
is only valid as long
as the pearl mass is much larger than the string mass, i.e., as long as
$$ b\left( {\tau \over b}\right) - \varphi> \left( {\lb \over b}\right)^{3}
{f^{6} \over \tau^{8}}
$$
For $f<\tau^2$ however part of the counterions have to condense onto the
outer pearl shell of thickness $\xi_{t}$ in order to decrease the surface
potential back to $\kt$.  In that case the results of the previous section
where charge regulation by the pearls was already considered apply.
Additionally we remark that the range of parameters is small for the validity of
these scaling arguments. Within the presented approach, we cannot guarantee
that the necklaces are stable. This remark does not at all affect the regimes
discussed, e.g., in \cite{lee:99.1}

\section{Conclusion}

A polyelectrolyte globule in a poor solvent has been shown by Kantor and
Kardar \cite{kantor:94,kantor:95} and Dobrynin and Rubinstein
\cite{dobrynin:99} to elongate with increasing charge and to have a necklace
structure comprising smaller "pearls" separated by stretched strings quite
similarly to the so-called Rayleigh instability of charged liquid droplets.
If we neglect the interaction with the charged monomers of the strings and of
the other drops,, the pearl size is fixed by a Rayleigh criterion, $l_B
Q^2/R\sim \tau R^2/b^2$. Most monomers are in pearls whereas the strings make
most of the chain length.  In this paper we discussed theoretically
pearl-necklace chain elasticity.

We first developed a continuous picture, valid for a necklace comprising many
pearls and where single pearl features are ignored.  When the necklace
elongates under the action of an external force, the pearls are unwinded and
converted into stretched strings. During this process the thermodynamic
equilibrium between pearl and string monomers is preserved; this imposes the
string tension to $\kt \tau/b$ and the string diameter, ($b/\tau)$. At
equilibrium the tension in a string balances the repulsion between
half-necklaces, and the external force. This force balance leads to the simple
force/length relation: $\varphi = \kt (\tau/b - \lb f^2N^2/L^2)$. During the
unwinding process the length increases by a factor $\Lambda^{1/2}$ where
$\Lambda>>1$ measures the number of thermal blobs in a pearl (or equivalently
$\Lambda$ is the inverse of the electrostatic interaction interaction between
thermal blobs). This is in contrast with the earlier cylindrical model which
predicts that when a force is applied to a spherical globule, the globule
becomes unstable when it is elongated by a factor of order two.

In practice, the number of pearls is not always very large (as seen in several
numerical simulations) and the discreteness of the pearls becomes an important
factor. We have discussed in details a necklace comprising two pearls.  When
the external force is increased, a well defined metastable single string state
appears in addition to the two pearl necklace.  For a somewhat higher force,
the single string and the two-pearl necklace exchange stability. At this
critical force, the pearl size is still close to the pearl size in the absence
of external force. When about half of the pearls is unwinded, the metastable
two-pearls state disappears and the single string state is the only
equilibrium conformation.  For a fast stretch-collapse cycle, a hysteresis
loop describing most of the metastable branches is expected. For quasi-static
cycles, a force plateau is anticipated at coexistence (due to the one
dimensional character of the problem the plateau is not associated to a real
phase transition).  For necklaces containing more pearls, our description uses
a somewhat simpler model, where the intensive string properties, tension
etc... are kept constant and where the only variables which are allowed to
fluctuate are the number of monomers in a pearl and the number of pearls; this
allows a transfer of the monomers from the pearls towards the strings as the
applied force is increased.  In this model the number of pearls is bound to be
an integer. In a first approximation neglecting thermal fluctuations, an
abrupt decrease of the pearl number by one unit is predicted for a given
length (sequence of lengths).  This formally corresponds to a abrupt decrease
of the force. There is thus a sequence of Maxwell loops, corresponding to
force plateaus in a quasi static transformation, or to a series of spinodal
length jumps under increasing applied force. This picture is altered by
fluctuations which are usually expected to be important and that smoothen the
plateaus.

The pearl-necklace structure of polyelectrolytes relies on a steepest descent
description which becomes exact in the thermodynamic limit for macroscopic
systems. In the usual Rayleigh instability of small (yet macroscopic) droplets
the energy scale, say typically the surface energy of a marginally stable
droplet, is large compared to $\kt$.  For mesoscopic systems such as the
polyelectrolyte necklace, this is not always the case and thermal fluctuations
are often important. In the case of a two pearl necklace we have computed the
average length , and the length fluctuation when the necklace is subject to an
external force. A pseudo-plateau is indeed found for the force but even for a
pearl self-energy as high as $100\kt$ the plateau is strongly rounded and the
length fluctuations in the plateau are roughly as large as the average length.
If the necklace contains many pearls, the fluctuations have also been taken
into account. The force is the sum of a continuous component that can be
calculated from the continuous model and a component arising from the
discreteness of the pearl number.  This last contribution is non monotonic and
can lead to Maxwell loops. However if the self-energy of a pearl is small or
if the network comprises many pearls the plateaus are strongly rounded by
fluctuations and significant force (pseudo) plateaus are no longer expected.

In the pearl-necklace structure, the electrostatic charge on the pearls is
rather high and, even though there is no Manning condensation on the strings
or on the smeared out necklace charge, condensation on the pearls (or charge
regulation) can occur.  In agreement with previous work, charge regulation is
found for $f<\tau^3$ and the polyelectrolyte collapses into a large globule
whenever counterion condensation becomes important. If an external force is
applied, the collapsed globule discontinuously stretches at a critical tension
of order $\kt\tau/b$.  The extreme case of a very poor solvent where there is
no ion dissociation in the (ionomer-like) dense phase has also been discussed
as it is of some experimental relevance. In this case stable necklaces are
obtained and the pearls unwind under an external force.

As a remark aside we would like to mention, that we ignored the effect of salt
which might be added to the solution. This will be part of a different
work. However, we expect that the results presented here are valid as long as
the persistence length (including the electrostatic contribution) together with
the Debye screening length is larger than $l$, i.e., the distance between two
pearls. 

Besides direct single chain elasticity measurements and implications for drag
reduction in polyelectrolyte solutions and related topics, the single chain
elasticity studied in this paper can be used as a basis to investigate more
complex systems.  In a separate work, we will present results on
polyelectrolyte gel elasticity in a poor solvent following Katchalsky's
\cite{katchalsky:51} arguments.

\subsection*{Acknowledgments} T.A.V. acknowledges the financial support of the
LEA. The hospitality of
the Institut Charles Sadron is gratefully appreciated. The authors
acknowledge enjoy- and helpful discussions with Phil
Pincus. They are also grateful to  H. Schiessel, who
sent his preprint \cite{schiessel:00.1} prior to publication.\\
\\
{\bf Note added:} After submission of this work, we received a
preprint by H. Schiessel \cite{schiessel:00.1}. In this work the same topic is
discussed. The free energies are discussed numerically and the results are
complementary to those discussed here.

\end{document}